\documentclass{amsart}  
\usepackage{geometry}  
\usepackage{graphicx}
\usepackage{amssymb}

\title{A direct dark matter detection experiment is inevitable}
\author{Vadim A. Bednyakov \\
{\em JINR, Joliot-Curie 6, 141980 Dubna, Moscow region, Russia. \lowercase{bedny@jinr.ru}}
}
\date{\today}  
\renewcommand{\baselinestretch}{1.2}

\begin{document} 
\begin{abstract}\normalsize     
This text contains the main message of my previous review \cite{Bednyakov:2015uoa} on the dark matter problem and supports resent paper \cite{Froborg:2020tdh}.  
True dark matter particles possess an exclusive galactic signature --- the annual modulation,  which is accessible today via direct dark matter detection only. 
One has no another way to prove the true nature of any dark matter candidate.		
\end{abstract}
\maketitle

\large

Dark Matter (DM) particles do not emit or reflect any detectable electromagnetic radiation 
and clearly manifest themselves today only gravitationally by affecting other astrophysical objects.
Numerous observational indications at astronomical and cosmological scales 
\cite{Zwicky:1933gu,Livio:2014gda,Drees:2012ji,Saab:2012th,Bertone:2004pz,Famaey:2015bba,Iocco:2015xga,Durazo:2015zzzz,McGaugh:2015tha,Iocco:2015bja,Sofue:2015xpa},
 	as well as results from very sophisticated numerical many-body simulations
	of the genesis of cosmic large- and small-scale structures (see, for example, 
\cite{Kuhlen:2012ft}), indicate the presence of this new form of matter in the Universe. 
 
	In particular, stars and gas clouds in galaxies and galaxies in clusters move 
	faster than can be explained by the pull of visible matter alone.
	Light from distant objects may be distorted by the gravity of intervening dark material.
	The pattern of the large-scale structures across the Universe is largely dictated by DM.
	In fact, about 85\% of the Universe's mass is dark, 
	accounting for about a quarter of the total cosmic energy budget
\cite{Livio:2014gda,Gelmini:2015zpa,Hoeneisen:2015rva}.	

	According to the estimates based on a detailed model of our Galaxy
\cite{Kamionkowski:1997xg}, the local density of DM amounts to about
$\rho_{\rm local}^{\rm DM} \simeq  0.4 - 0.5 {\rm ~GeV/cm}^3 $
\cite{Famaey:2015bba,Pato:2015dua}. 
	The local flux of DM particles, which can cross 1\,cm$^2$ of the Earth's surface 
	each second contains about $10^5$ DM particles, 
	provided their mass is 100 GeV/$c^2$.
	This value is often considered as a promising basis for 
	laboratory direct DM search experiments.

	The Big Bang conception of the early Universe
\cite{Kolb:1990vq} strongly supports the idea of non-gravitational coupling 
	of DM particles with ordinary matter.
	This interaction could be very weak, but not completely vanishing. 
	 
	Despite many other possibilities
\cite{Feng:2010gw}, the Weakly Interacting Massive Particle (WIMP) is among the
	most popular candidates for the relic DM.
	Being electrically neutral and interacting rather weakly, WIMPs 
	naturally reproduce the correct relic DM abundance. 
	These particles are non-baryonic,  and there is no room for them in the 
	Standard Model of particle physics (SM), 
	in particular due to the Big Bang nucleosynthesis, which successfully predicts 
	abundances of light elements such as deuterium, helium and lithium 
	arising from interactions in the early Universe.
	Furthermore, to explain the way in which galaxies form and cluster, 
	these massive DM particles should be non-relativistic, or so-called "cold DM" particles.

	The primary goal of modern particle physics and astrophysics is to detect 
	the DM particles that constitute the massive invisible halo of the Milky Way.  This "DM Problem" is a real challenge for modern physics and 
	experimental technology.
	To solve the problem, i.e., {\em at least}\/ to detect these DM particles, 
	one simultaneously needs to apply the front-end knowledge 
	of modern particle physics, astrophysics, cosmology and nuclear physics.
	Furthermore,  one should develop and have in long-term usage an 
	extremely highly sensitive setup, to say nothing about 
	complex data analysis methods.

	The DM problem can be solved 
	by means of a complete and balanced research program based on 
	the following four categories. \\
\underline{Direct Detection experiments} look for a direct DM interaction in an underground terrestrial 
	low-background laboratory, where a DM particle scatters off a (nuclear) substance of a detector, 
	producing a detectable recoil and/or ionization signal.  \\
\underline{Indirect Detection experiments} are unable to detect DM directly,
	 but, using (huge) terrestrial setups, try to detect products of DM particle annihilations 
	 in cosmic objects like the Earth, the Sun, our own and/or another galaxy.
	It is assumed that pairs of DM particles annihilate each other producing high-energy 
	ordinary particles (antimatter, neutrinos, photons, etc).  
	In some models, the DM particles can be metastable and eventually decay 
	with the production of SM particles. \\
\underline{Particle Collider experiments}  can help one to understand the properties of the DM particles. 
	The LHC and future lepton and hadron colliders 
	can produce energetic DM particles that will obviously escape detection 
	but could be registered by means of an excess of events with missing energy or momentum. \\
\underline{Astrophysical Probes} provide information about
	non-gravitational interactions of DM particle populations, 
	such as DM densities in the centers of galaxies and cooling regimes of stars.
	The particle properties of DM are constrained here through observation of 
	their joint impact on astrophysical observables.  

\begin{figure}[!ht] 
\includegraphics[width=\columnwidth]{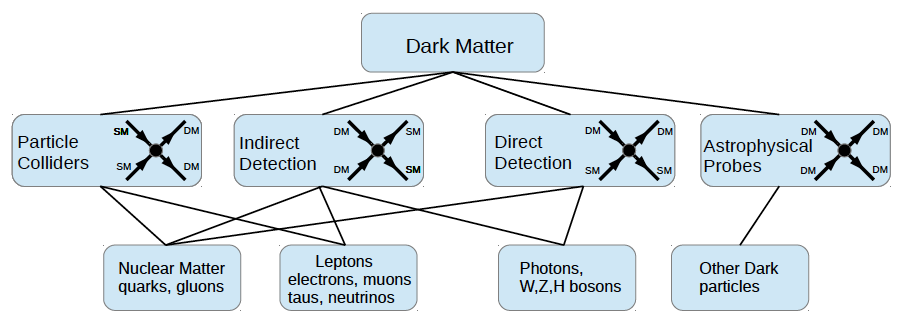}
\caption{The idea of DM particle registration with a terrestrial setup 
	relies on the common belief that DM can interact non-gravitationally with 
	nuclear matter, leptons, gauge and other bosons, and with other possible dark particles.  
	These interactions may be studied nowadays with
  	direct and indirect detection techniques, particle colliders, and via astrophysical probes 
 \cite{Bauer:2013ihz}. } \label{fig:2013ihz-DM-interplay}
\end{figure}

	These search strategies are schematically shown in 
Fig.~\ref{fig:2013ihz-DM-interplay}. 
	
	In order to convincingly establish the nature of a DM candidate,
	one must reach consistency between all possible DM searches  
	for the common DM candidate parameters of mass, spin, and coupling strengths
\cite{Christensen:2014yya}.
 
	 The direct DM detection has an exceptional 
	 status among the other DM search techniques discussed above. 
	 The reasons could be a rather old history of this approach, 
	 existence of the DAMA evidence
\cite{Bernabei:2013xsa},
	  and a possibility of supplying us with the clearer and most decisive information on the DM problem.	 
	
	Furthermore,
	only direct DM observation can prove the existence of DM particles.
	Only in a direct DM experiment one can have a chance to see 
	the galactic nature of the true DM particle population via
	measuring the annual modulation of the recoil signal.
	This signature is utterly needed to prove DM existence. 
	There is no way to ignore the key role of direct DM detection.
	The other DM search approaches --- indirect, collider, astrophysical --- 
	can only assist the direct DM detection, for example, 
	with mass region search advice, local relic density estimates, etc. 
	 
\renewcommand{\baselinestretch}{1.18}
\providecommand{\href}[2]{#2}\begingroup\raggedright\endgroup
 
\end{document}